\documentstyle[referee]{mn}
\newcommand{\reference}{\bibitem}
\newcommand{\beq}{\begin{equation}}              
\newcommand{\beqa}{\begin{eqnarray}}             
\newcommand{\eeq}{\end{equation}}                
\newcommand{\eeqa}{\end{eqnarray}}               
\newcommand{\eeqi}[1]{\quad#1\end{equation}}     
\newcommand{\eeqai}[1]{\quad#1\end{eqnarray}}    

\newcommand{\apj}{ApJ}
\newcommand{\aj}{AJ}
\newcommand{\mnras}{MNRAS}
\newcommand{\eps}{\epsilon}
\newcommand{\phis}{\phi_{\rm s}}
\def\lesssim{\mathrel{\hbox{\rlap{\hbox{\lower4pt\hbox{$\sim$}}}\hbox{$<$}}}}
\def\gtrsim{\mathrel{\hbox{\rlap{\hbox{\lower4pt\hbox{$\sim$}}}\hbox{$>$}}}}

\input epsf

\title[]{On the Magnfication Relations in Quadruple Lenses: A Moment Approach}

\author[Witt \& Mao]
{
Hans J. Witt$^1$\thanks{e-mail: hwitt@aip.de} 
Shude Mao$^2$\thanks{e-mail: smao@mpa-garching.mpg.de} \\
$^1$Astrophysikalisches Institut Potsdam, An der Sternwarte 16, 
        14482 Potsdam, Germany \\
$^2$Max-Planck-Institute f\"ur Astrophysik,
        Karl-Schwarzschild-Strasse 1, 85740 Garching, Germany}

\date{Accepted ........
      Received .......;
      in original form .......}
\pagerange{\pageref{firstpage}--\pageref{lastpage}}
\pubyear{1998}

\begin{document}
\maketitle
\label{firstpage}

\begin{abstract}
We present a new method of studying quadruple lenses in elliptical
power-law potentials parameterized by
$\psi(x,y) \propto (x^2+y^2/q^2)^{\beta/2}/\beta~
(0 \leq \beta < 2)$.  For this potential,
the moments of the four image positions weighted by signed magnifications 
(magnification times parity) have very simple properties.
In particular, we find that the zeroth moment -- the
sum of four signed magnifications 
satisfies $\simeq 2/(2-\beta)$; the relation is exact for 
$\beta=0$ (point-lens) and $\beta=1$ (isothermal potential),
independent of the axial ratio.
Similar relations can be derived when a shear is present along the major
or minor axes. These relations, however, do not hold well for 
the closely-related elliptical density distributions.
For a singular isothermal elliptical density distribution without
shear, the sum of signed magnifications for quadruple lenses
is $\approx 2.8$, again nearly independent of the ellipticity. For the
same distribution with shear, the total signed
magnification is around 2-3 for most cases, but can be significantly
different for some combinations of the axial ratio and shear where 
more than four images can appear.
\end{abstract}

\section{INTRODUCTION}

Since the discovery of the first double lens 0957+0561 by
Walsh, Carswell \& Weymann (1979),
gravitational lensing has found many cosmological applications
(see Schneider, Ehlers \& Falco 1992; Narayan \& Bartelmann 1998 and
references therein). One of these applications is to probe the potential
of lensing galaxies in multiply-imaged quasars. In this aspect, the
quadruple lenses are ideal systems since they provide more constraints
than the double lenses. Previous modelings of these systems use mainly
isothermal elliptical potentials and their variants (e.g., Kochanek 1991;
Keeton, Kochanek \& Seljak 1997; Witt, Mao \& Schechter 1995). 
In many of these modelings,
attention was paid only to the image positions with a few exceptions,
which considered the flux ratios (e.g., Keeton,
Kochanek \& Seljak 1997).
Since optical flux ratios are likely to be influenced by
microlensing (Chang \& Refsdal 1979),
they should be used with care. On the other hand, radio fluxes should be
nearly unaffected by microlensing since the radio emission regions
are believed to have larger
sizes (there are exceptions for very highly magnified
systems such as B1422+231, see Mao \& Schneider 1998). Satisfactory lens
models should ideally fit the image positions as well as the flux
ratios. However, previous numerical studies often find it
very difficult to fit these with isothermal potentials and their
variants (e.g., Keeton, Kochanek \& Seljak 1997).
The purpose of this paper is two-
fold: first we extend the isothermal potentials to a more general class of
power-law potentials and uncover some generic scaling relations.
The three-dimensional correspondence of this class
of potential provide a reasonable analytical approximation to the galactic
potential (Evans 1994). Second, we study the relations for related
isothermal elliptical density distributions and show that
there are qualitative and quantitative differences
between elliptical density and elliptical potential distributions.

The outline of the paper is as follows. In \S 2, we introduce the elliptical
power-law potentials and derive some scaling relations with a new
moment approach. In \S 3, we study the singular isothermal elliptical density
distribution  with arbitrary shear
to understand the difference between these two types of
potentials. In \S 4, we summarize and discuss the implications of our 
results for lens modeling.

\section{ELLIPTICAL POWER-LAW POTENTIAL}

In this section we study elliptical power-law potentials
as models for lensing galaxies.
This potential form is widely used in gravitational lensing
(e.g., Blandford \& Kochanek 1987;
Kassiola \& Kovner 1993 and references therein). It is easier
to study than the more realistic but more complex
elliptical density distribution (e.g., Kormann, Schneider \&
Bartelmann 1994; Hogg \& Blandford 1994;
Chae, Khersonsky, \& Turnshek 1998); we return to
the elliptical density distributions in \S 3.

The power-law potential is given by
\beqa \label{phi}
\psi(x,y) &=&  
a ~\frac{(x^2+y^2/q^2)^{\beta/2} }{ \beta } 
\quad \mbox{for} \quad  0 < \beta < 2  \\
\psi(x,y) &=&  
a ~\frac{\ln (x^2+y^2/q^2)}{2}
\quad \mbox{for} \quad  \beta = 0
\eeqa
where $q$ ($0<q\le 1$)
is the axial ratio of the lensing galaxy, $\beta$ the slope of 
the potential and $a$ determines the overall strength of the lens system. 
The coordinate system has been chosen such that the center of the lensing
galaxy is at the origin and the major axis of the galaxy is along the
$x$-axis. In general the orientation of the major axis of the lensing
galaxy is unknown, but can be easily
determined if the image configurations can be 
modeled by an elliptical potential
(Witt 1996). Note that the
point-mass potential is described by $\beta=0, q=1$, the 
elliptical isothermal sphere by $\beta=1$, and the 
constant surface mass distribution  by $\beta \rightarrow 2, q=1$.
Thus the commonly used potentials in numerical modeling are all
special cases of this class.

\subsection{Lens equation}

The lens equation for the power-law potential is given by
\beqa 
\xi  &=& x -\frac{a x}{(x^2+y^2/q^2)^{1-\beta/2} }, \label{xis} \\ 
\eta &=& y -\frac{a y q^{-2}}{(x^2+y^2/q^2)^{1-\beta/2} }. \label{etas} 
\eeqa
It is easy to show that the lens equation with an {\it on-axis} shear
can be transformed into eqs. (\ref{xis}) and (\ref{etas})
with suitable coordinate transformations as given in \S 2.5.
So with small modifications
our results apply to the case with an on-axis shear as well
(\S 2.5). Notice that
the elliptical potentials (including the special power-law form studied here)
restrict the image and galaxy positions to be
located on a hyperbolic line (Witt 1996). Image configurations which
do not follow these constraints require far more complicated lens models,
in particular may require an off-axis shear (Keeton, Kochanek \& Seljak 1997;
Witt \& Mao 1997).

\subsection{Exact relations for $\beta=1$ and $\beta=0$}

Although the lens equations (\ref{xis}) and (\ref{etas})
are in general too complicated to solve
for the image positions and the magnifications,
here we present a new method to directly relate
the image magnification and positions to the properties
of the galaxy potential. The new method is similar to the Jeans
equation in galactic dynamics (e.g., Binney \& Tremaine 1987,
p. 195). It is possible to extend this new technique to other
potential forms such as that for binary lens.
However, in the latter case, 
the application may be limited since in microlensing 
we can not resolve the individual images.
In the next subsection, we illustrate our technique using 
an isothermal ($\beta=1$) elliptical potential; in \S 2.3, we discuss
the case for $\beta=0$.

\subsubsection{Isothermal elliptical potential ($\beta=1$)}

There are different ways to derive relationships between the 
image positions and their magnifications. Sometimes it is more convenient
to use the complex formalism (cf. Rhie 1997 for the binary lens application)
or the resultant method (e.g., Witt \& Mao 1995; Dalal 1998).
Here we adopt a general analytical approach using real quantities.
When a source is inside the inner caustic of an elliptical potential,
there are always four images. The lens equation 
(\ref{xis}) or (\ref{etas}) can be manipulated into
two polynomials of degree 4 of separated arguments in $x$ and $y$:
\begin{equation} \label{g1-g2}
g_x(x,\xi,\eta)= \sum_{n=0}^{4} a_n x^n =
a_{4} \prod_{i=1}^{4} (x-x_{i}),~~~
g_y(y,\xi,\eta)= \sum_{n=0}^{4} b_n y^n =
b_{4} \prod_{i=1}^{4} (y-y_{i}) 
\end{equation}
where $(x_i, y_i)$ are the coordinates of the $i$-th image, and
the polynomial coefficients are given by
\beqa
a_4 &=& (1-q^2)^2, \label{a4eq}  \qquad    
b_4  =  q^2(1-q^2)^2, \\
a_3 &=& 2 \xi (1-q^2) (q^2-2), \quad  
b_3  =  2 \eta q^2 (1-q^2) (2q^2-1), \\
a_2 &=& -a^2 (1-q^2)^2 +6 \xi^2 (1-q^2) +q^2 (q^2 \xi^2 +\eta^2) \quad
b_2  = a_2 -6 (\eta^2 q^4+\xi^2) (1-q^2)   \\
a_1 &=&  2 \xi [ a^2 (1-q^2) - 2 \xi^2 + q^2 (\xi^2 -\eta^2)[     \quad
b_1  =  - 2 q^2 \eta [ a^2 (1-q^2) + 2 q^4 \eta^2 + q^2 (\xi^2 -\eta^2)] \\
a_0 &=& - \xi^2 [a^2 -  (\xi^2 +q^2 \eta^2)], \quad 
b_0 = - \eta^2 q^4 [a^2 - (\xi^2 +q^2 \eta^2)]  \label{b0eq}
\eeqa
The Jacobian matrix for the mapping
from the lens plane to the image plane is defined as
\begin{equation}  \label{Jacobian}
J = 
\left( \matrix{ {\partial \xi \over \partial x} & 
{\partial \xi \over \partial y } \cr
{\partial \eta \over \partial x } & 
{\partial \eta \over \partial y } \cr } \right) 
\end{equation}
For each image, there is a one-to-one inverse mapping
from the image plane to the lens plane,
\begin{equation}  \label{invJ}
J^{-1} = 
\left( \matrix{ {\partial x \over \partial \xi} & 
{\partial x \over \partial \eta } \cr
{\partial y \over \partial \xi } & 
{\partial y \over \partial \eta } \cr } \right) 
= \frac{1}{\det J}
\left( \matrix{ {\partial \eta \over \partial y} & 
-{\partial \xi \over \partial y } \cr
-{\partial \eta \over \partial x } & 
{\partial \xi \over \partial x } \cr } \right),
\end{equation}
where the second equality comes from the conventional definition of
an inverse matrix. A comparison of the diagonal elements gives
\beq \label{diag}
{\partial y \over \partial \eta } = \frac{1}{\det J} ~~
{\partial \xi \over \partial x },~~
{\partial x \over \partial \xi } =  \frac{1}{\det J} ~~
{\partial \eta \over \partial y }
\eeq
The sum of the two expressions is
\beq \label{key1}
{\partial y \over \partial \eta } +  
{\partial x \over \partial \xi }  = \frac{1}{\det J}
\left[ {\partial \xi \over \partial x} + 
{\partial \eta \over \partial y } \right] = \frac{1}{\det J} +1,
\eeq
where we have used 
\beq
{\partial \xi \over \partial x} + 
{\partial \eta \over \partial y }  = 1+ \det J ~~
{\rm for ~~ \beta=1}.
\eeq
The signed magnification for a given image is by definition
\begin{equation} \label{mup}
p_i \mu_i \equiv
\left. \frac{1}{\det J} \right\vert_{(x_{i},y_{i})}
= {\partial y_i \over \partial \eta } +  
{\partial x_i \over \partial \xi } -1,
\end{equation}
where $\mu_i$ is the absolute magnification for a given image, 
$p_i=\pm 1$ is the parity of of the images, and we substituted
$\det J$ from eq. (\ref{key1}).

Now the sum of signed magnifications is then
\begin{equation} \label{sumpmu}
\sum_{i=1}^{4} p_i \mu_i =
\frac{\partial}{\partial \xi} \sum_{i=1}^{4} x_{i}(
\xi,\eta)+ 
\frac{\partial}{\partial \eta} \sum_{i=1}^{4} y_{i}(
\xi,\eta)   -4
= 
\frac{\partial}{\partial \xi} \left(-\frac{a_3}{a_4}
\right)
+\frac{\partial}{\partial \eta} \left(-\frac{b_3}{b_4}
\right) -4.
 \end{equation}
In the second step
we have used the fact that the sum of the roots of a polynomial
is related to its coefficients. Using the coefficients as given
in eqs. (\ref{a4eq}) to (\ref{b0eq}), we obtain
\begin{equation}
\sum_{i=1}^{4} p_i \mu_i = 2.
\end{equation}
A similar relation was found recently by Dalal (1998) for $\beta=1$.
This implies that for an isothermal elliptical potential the signed
magnification is always 2, independent of the axial ratio
and the image configurations.

Using similar techniques, we can 
calculate higher order moments of positions, defined as
\beq
{\cal M}_{n,x} \equiv \sum_{i=1}^{4} p_i \mu_i x_{i}^n,  ~~
{\cal M}_{n,y} \equiv \sum_{i=1}^{4} p_i \mu_i y_{i}^n 
\eeq
For illustration, we derive the $x$-component of the first moment:
\begin{equation}
{\cal M}_{1, x} = \sum_{i=1}^{4} 
p_i \mu_i x_{i} = 
\sum_{i=1}^{4} \left[
{\partial y_{i} \over \partial \eta } +  
{\partial x_{i} \over \partial \xi } -1
\right] x_{i},  
\end{equation}
where in the second equality
we have again used eq. (\ref{key1}). Changing now the order of 
summation and derivation, and 
using the fact that the Jacobian matrix is symmetric 
($\partial y /\partial \xi =\partial x/\partial\eta$), we obtain:
\beq \label{mix}
\sum_{i=1}^{4} p_i \mu_i x_{i} = 
\frac{1}{2} \frac{\partial}{\partial \xi} \sum_{i=1}^{4} (x_{i}^2 -y_{i}^2)
+ \frac{\partial}{\partial \eta} \sum_{i=1}^{4} x_{i} y_{i} 
- \sum_{i=1}^{4}  x_{i}
\eeq
For the first and third terms we can easily replace the sum
by expressions of the coefficients of the polynomial:
\beq
\sum_{i=1}^{4} x_{i}^2 = \frac{a_3^2 - 2 a_2 a_4}{a_4^2},
\quad 
\sum_{i=1}^{4} y_{i}^2 = \frac{b_3^2 - 2 b_2 b_4}{b_4^2},
\quad \mbox{and} \quad \sum_{i=1}^{4} x_{i}= - \frac{a_3}{a_4}.
\eeq
For the second mixed term in $x$ and $y$ of eq. (\ref{mix}),
we recall that for any elliptical
potential, the $x$ and $y$ components for a given image are related
(cf. eqs. [\ref{xis}] and [\ref{etas}]) by:
\beq \label{xyrel}
\frac{\xi - x_i}{\eta - y_i} = q^2 \frac{x_i}{y_i}, \quad {\rm thus} \quad 
y_{i} = \frac{\eta q^2 x_{i}}{ \xi - x_{i} (1-q^2)}
\eeq
Now the mixed term can be written as
\beqa
\sum_{i=1}^{4} 
x_{i} y_{i} &=&  \frac{\eta q^2}{1-q^2}
\sum_{i=1}^{4} \frac{x_{i}^2}{ \xi /(1-q^2) - x_{i}} 
\nonumber \\ &=&
\frac{\eta q^2}{1-q^2} \left[ \frac{a_3}{a_4} - \frac{4 \xi}{1-q^2}
+ \frac{\xi^2}{(1-q^2)^2} \sum_{i=1}^{4} 
\frac{1}{ \xi /(1-q^2) - x_{i}} \right]
\eeqa
where we have expressed the mixed term in terms of a sum of different powers
of $x_{i}$.
The last term in the previous equation can be expressed in terms
of the polynomial:
\beq
\sum_{i=1}^{4} 
\frac{1}{ x - x_{i}} 
\left.\right|_{x=\xi/(1-q^2)}=
\frac{1}{g_x(x,\xi,\eta)} \frac{d g_x(x,\xi,\eta)}{d x}
 = {2(1-q^2) \over \xi q^2}
\eeq
Collecting all terms together, we obtain
\beq
\sum_{i=1}^{4} p_i \mu_i x_{i} = 2 \xi, ~~~
\sum_{i=1}^{4} p_i \mu_i y_{i} = 2 \eta, 
\eeq
a remarkably simple result.
Higher order moments can be 
obtained similarly and are summarized in \S 2.4. However, first we
discuss results for another special case with $\beta=0$.

\subsection{Point lens $(\beta=0)$}

For $\beta = 0$, the potential approximates a point lens plus shear.
It is easy to separate the lens equations into 
two polynomials of degree 4 for $x$ and $y$ -- We only have to insert
eq. (\ref{xyrel}) into the lens equations (\ref{xis}) and (\ref{etas}).
Therefore we do not
present the coefficients here.
The derivations for magnification moments are similar, except that
eq. (\ref{key1}) must be changed. To do this, we notice
that for $\beta=0$, we have
\beq
{\partial \xi \over \partial x} + q^2 
{\partial \eta \over \partial y }  = 1+ q^2.
\eeq
Using eq. (\ref{diag}), we obtain
\beq
 {\partial y \over \partial \eta } + q^2 
{\partial x \over \partial \xi }  = \frac{1}{\det J}
\left[ {\partial \xi \over \partial x} + q^2 
{\partial \eta \over \partial y } \right] = \frac{1+q^2}{\det J}.
\eeq
In the last step, we have
expressed the signed magnification ($\mu_i p_i = 1/\det J$) in
terms of ${\partial y / \partial \eta }$ and
${\partial x / \partial \xi }$, similar to the case for $\beta=1$.
The rest of derivation proceeds nearly identically. For the zeroth
and first moments we have
\beq \label{mubeta0} 
\sum_{i=1}^4 \mu_i p_i = 1,~~
\sum_{i=1}^{4} p_i \mu_i x_{i} = \xi, ~~~
\sum_{i=1}^{4} p_i \mu_i y_{i} = \eta, 
\eeq

\subsection{Relations for more general power-law potentials}

For more general cases with non-integer values between 0 and 2, there is
no simple way to derive analogous relations. However, the special cases
for $\beta=0$ and $\beta=1$ suggests approximation by
interpolation. In particular, 
\beq \label{mubeta} 
\sum_{i=1}^4 \mu_i p_i \simeq B, ~~B \equiv \frac{2}{2-\beta}.
\eeq
For $\beta=0$ and $\beta=1$, eq. (\ref{mubeta}) is exact.
For other non-integer values, this equation is only approximate, 
but we find numerically that it is valid to $\la$5\%. Therefore,
for an quadruple lens well-modeled by power-law elliptical potentials,
this relation is approximately valid, independent of the source position.
A similar exact relation is also satisfied by the binary lens, where the
sum of signed magnifications of five image cases is equal to 2
(Witt \& Mao 1995; Rhie 1997). 

Starting from eq. (\ref{mubeta}), we can obtain moment equations
that relate the image positions and magnifications
to the parameters of the potential.
\beqa \label{m1}
\sum_{i=1}^4 \mu_i p_i x_i  &=&   B \xi, 
~~ \sum_{i=1}^4 \mu_i p_i y_i  =   B \eta, \\
\label{m2} 
\sum_{i=1}^4 \mu_i p_i x_i^2  &=&   
- A + B \xi^2, 
~~ \sum_{i=1}^4 \mu_i p_i y_i^2 = A q^{-2 B}
+ B \eta^2, \\ 
\label{m3}
\sum_{i=1}^4 \mu_i p_i x_i^3 &=& 
- A \xi (B + \frac{2-q^2}{1-q^2})
+ B \xi^3,
~~
\sum_{i=1}^4 \mu_i p_i y_i^3 =
A \eta q^{-2 B}
(B + \frac{1-2q^2}{1-q^2}) + B \eta^3, 
\eeqa
where
$
A = a^B B q^2/(1-q^2).
$
Similar to eq. (\ref{mubeta}), these relations are exact only for
$\beta=0$ and $\beta=1$. However, numerical tests show that
these formulae are accurate to $\la$1\%
for other non-integer values of $\beta$ (or $B$).

\subsection{The inclusion of an on-axis shear}

The lens equation with an on-axis shear is given by
\beqa 
\xi  &=& x +\gamma x-\frac{a x}{(x^2+y^2/q^2)^{1-\beta/2} } \\
\eta &=& y -\gamma y-\frac{a y q^{-2}}{(x^2+y^2/q^2)^{1-\beta/2} } 
\eeqa
We can show that these equations can be transformed into 
equations (\ref{xis}) and (\ref{etas}) by a simple
transformation. It can be easily verified that the transformation
is given by
\beqa \label{xi-tran}
\xi' &=& \frac{\xi}{\sqrt{1+\gamma}} \qquad  
\eta' = \frac{\eta}{\sqrt{1-\gamma}} \\  
x' &=& x \sqrt{1+\gamma} \qquad y' = y \sqrt{1-\gamma} \\
a' &=& a (1+\gamma)^{-\beta / 2} \quad \beta' = \beta \quad
q' = q \sqrt{\frac{1-\gamma}{1+\gamma}},
\eeqa  
where we denote the transformed quantities with a prime. Applying the
above transformation, we obtain
\beqa 
\xi'  &=& x' -\frac{a' x'}{(x'^2+y'^2/q'^2)^{1-\beta'/2} } \label{xiprime}, \\
\eta' &=& y' -\frac{a' y' q'^{-2}}{(x'^2+y'^2/q'^2)^{1-\beta'/2} }.
\label{etaprime} 
\eeqa
These are identical to eqs. (\ref{xis}) and (\ref{etas}) if we drop the
prime. 

The magnification transforms like
\beq \label{mu-tran}
\mu_i' = \mu_i (1-\gamma^2)
\eeq

Now it is straightforward
to generalize the moment relations (eqs. [\ref{mubeta}-\ref{m3}])
to the case with an on-axis shear by applying
eqs. (\ref{xi-tran})-(\ref{mu-tran}). In particular, for the zeroth and
first moments we have 
\beq \label{mumu}
\sum_{i=1}^4 \mu_i p_i =  \frac{2}{(2-\beta)(1-\gamma^2)},
\eeq
\beq
\sum_{i=1}^4 \mu_i p_i x_i =  \frac{2\xi}{(2-\beta)(1-\gamma^2)(1+\gamma)},~~
\sum_{i=1}^4 \mu_i p_i y_i =  \frac{2\eta}{(2-\beta)(1-\gamma^2)(1-\gamma)},
\eeq
Notice that 
the inclusion of an on-axis shear changes the magnification relation
(eq. [\ref{mumu}]) only {\it quadratically}.
It is straightforward but tedious to show that this relation holds even for an
off-axis shear (cf. Dalal 1998).
For the observed gravitational lenses, the external shear is 
usually between $0<\gamma <0.3$ (Keeton, Kochanek \& Seljak 1997),
this would mean that for given
slope of the elliptical power-law potential,
the total sum of signed magnifications is known to within 10\%.

Two interesting observations are worth making here. First,
these simple formulae indicate that the moments of $x$ ($y$) components
${\cal M}_{x,n}$ of the images are only related to the $\xi$ ($\eta$)
component of the
source position. This seems remarkable since the magnification is in
general a function of both $\xi$ and $\eta$. Second, for an
elliptical potential with an on-axis shear, we have 7 unknown quantities
($\xi,\eta,a,q,\beta,\mu_1, \gamma$) with 7 moment equations, 
so in principle we can use the moments to solve for the lens parameters
directly. Indeed we have implemented such a routine. Unfortunately,
using Monte Carlo simulations, we find that the results
depend rather sensitively on the accuracy of image positions and
flux ratios. The present observational uncertainties do not allow us
to carry out such an exercise, although in the future such 
an effort may be more worthwhile in order to determine whether elliptical
potentials provide adequate fits to the observed lenses.
In any case, we will show next that the magnification relations
do not hold exactly for the elliptical density distributions and
therefore these relations should be used with some caution.

\section{Singular Isothermal Elliptical Density Distributions}

It is well known that power-law potentials develop unphysical features,
such as dumb-bell shaped density contours or even negative densities,
when the axial ratio is small (e.g., Kassiola \& Kovner 1993).
So it is important to check whether the relations derived
for the simpler elliptical potentials are valid for the more realistic 
elliptical density distributions. Relations concerning only image
positions seem to be equally well satisfied by the elliptical potentials
and elliptical density distributions (Witt \& Mao 1997).
However, this is not necessarily true
for the magnification relations since
magnifications involve second order
derivatives of the lensing potential rather than the first order for the image
positions, and hence they could be
more sensitive to the detailed potential forms. We show that
indeed elliptical density and elliptical potentials have somewhat
different magnification relations. In this paper,
we focus on the singular elliptical density distribution
(including an arbitrary shear) for two reasons: the first
is that this model can reproduce the observed image configurations well,
particularly the models incorporating
a shear (Keeton, Kochanek \& Seljak 1997).
The second reason is that relatively
little is known about the magnification properties for the elliptical
density distributions, so a complete treatment of the general case is
beyond the scope of this paper. 

In polar coordinates ($r, \theta$), the surface
density for the singular isothermal distribution is given by
\beq \label{sed}
\kappa(r, \theta) = {a\over 2r}{1 \over (1-\eps \cos 2\theta)^{1/2}},
\eeq
where $\epsilon=(1-q^2)/(1+q^2)$, $q$ is the axial ratio
and $a$ is again a characteristic
radius that determines the overall image separations.
For $\eps>0$, the major axis is along the $x$-axis. The
lens equation (including a shear) is 
\beqa
\xi &= &(1+\gamma_1) x + \gamma_2 y- {a\over \sqrt{2\eps}} \tan^{-1}
\left[ \sqrt{2\eps} \cos\theta \over (1-\eps \cos 2\theta)^{1/2} \right],
\\
\eta &=& \gamma_2 x + (1-\gamma_1)y - {a\over \sqrt{2\eps}} \tanh^{-1}
\left[ \sqrt{2\eps} \sin\theta \over (1-\eps \cos 2\theta)^{1/2} \right],
\eeqa
where 
$(\gamma_1=\gamma \cos 2\phis,\gamma_2=\gamma \sin 2\phis$
with $\gamma$ being the strength of shear and $\phis$ the
``direction'' of shear. The magnification for a given image is
\beq
(\det J)^{-1}  = 1 - 2\kappa (1+\gamma_1 \cos 2\theta + \gamma_2 \sin
2\theta) -\gamma^2.
\eeq
These results can be found in Kormann et al. (1994) and Keeton \&
Kochanek (1998). Below we first discuss the simpler case without
shear and then discuss some examples of the
general case with shear (in an arbitrary direction).

\subsection{The case without shear ($\gamma=0$)}

For this case, we have four (two) images if the source is inside (outside)
a diamond-shaped caustic (see Fig. 1). Again, we study the sum of signed
magnifications for the four image case. Recall that for an elliptical
potential with the same power-law slope ($\beta=1$), the sum of signed
magnification is exactly 2. Now for the isothermal elliptical density
distribution, surprisingly, we find a value very close to 2.8,
{\it nearly independent of the ellipticity}. For axial ratio in the range of
$0.1<q<1$, the sum of signed magnification varies little, from 2.65 to 2.8.
This can be demonstrated by considering the special case where the
source is located at the origin. The lens equation can be easily solved
and the total signed magnification is given by
\beq
\sum_{i=1}^{4} \mu_i p_i = {2 \over 1- u/\tan^{-1} u}
+{2 \over 1 - v/\tanh^{-1} v},
\eeq
where $u=[2\eps /(1 - \eps)]^{1/2}$ and $
v=[2\eps /(1 + \eps)]^{1/2}$. The above expression can be expanded
in series around $\eps=0$, 
\beq
\sum_{i=1}^{4} \mu_i p_i =
{14 \over 5}-{32 \over 875} \eps^2 -{490272 \over 21896875} \eps^4 + O[\eps^6]
\eeq

The expansion only involves even powers of $\eps$ and 
the coefficients are very small, which 
explain why the sum is nearly independent of $\eps$ (or $q$). Numerically, we
find that the sum varies $\leq 10\%$ as long as we have four images,
i.e.,  when the source is inside the caustics.
To illustrate the near constancy of the sum of signed magnifications,
we show in Fig. 1, the magnification contours for two values of $q$:
a nearly spherical case with $q=0.9$ and a fairly elongated case with
$q=0.3$; these values bracket the observed light axial ratios in early types
of galaxies. 
The contour shapes are qualitatively the same; the case with $q=0.9$
has a much smaller range in the total signed
magnification (varies from $\approx 2.79$ to $\approx 2.82$)
than the more elongated case. We emphasize that 
although the sum is nearly constant, it is 
quite different from the value (2) of the isothermal potential. This
cautions against simple extrapolation of the magnification 
results obtained 
from elliptical potentials to elliptical density distributions.

\def\plottwo#1#2{\centering \leavevmode
\epsfxsize=.45\columnwidth \epsfbox{#1} \hfil
\epsfxsize=.45\columnwidth \epsfbox{#2}}

\begin{figure}
\plottwo{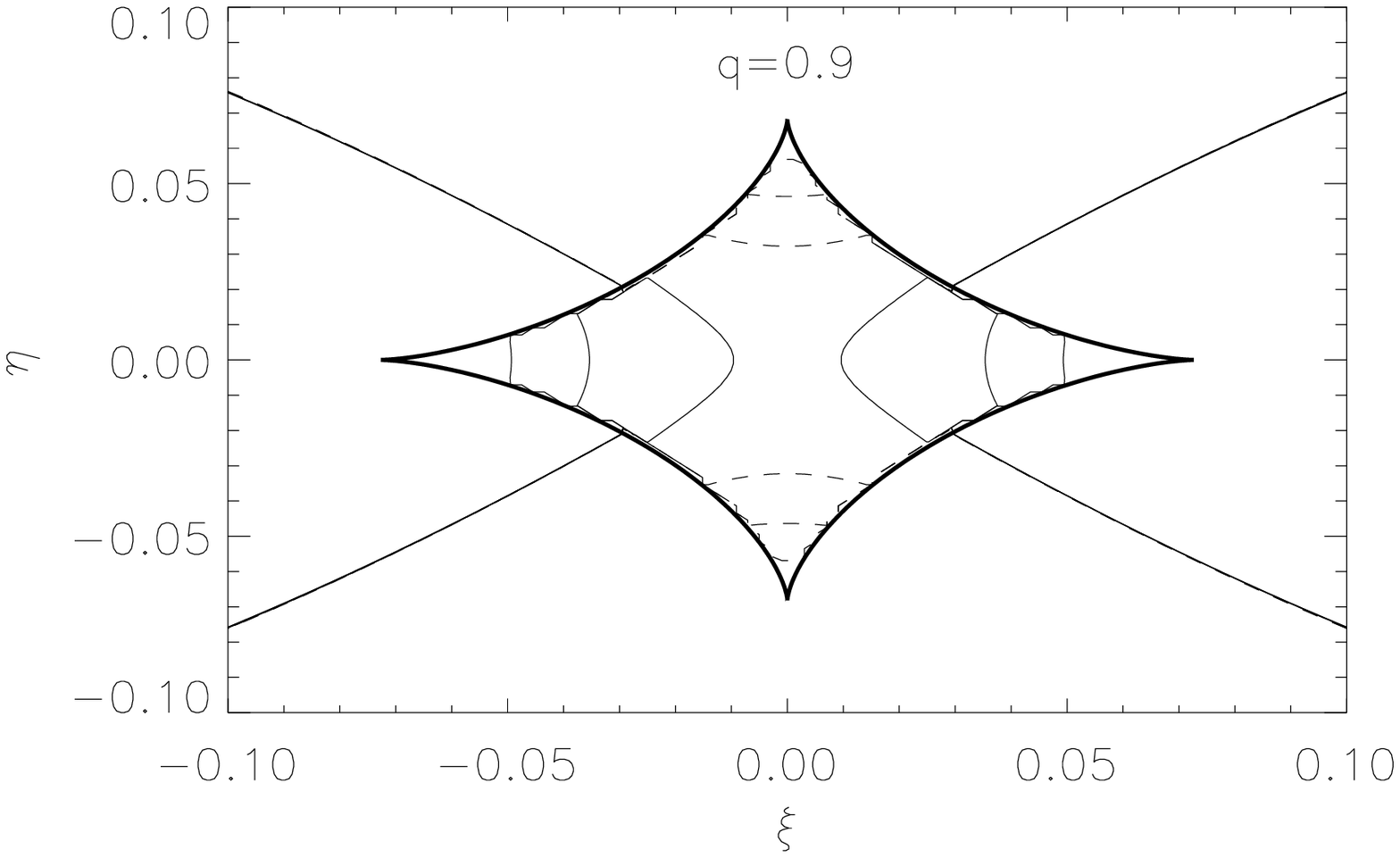}{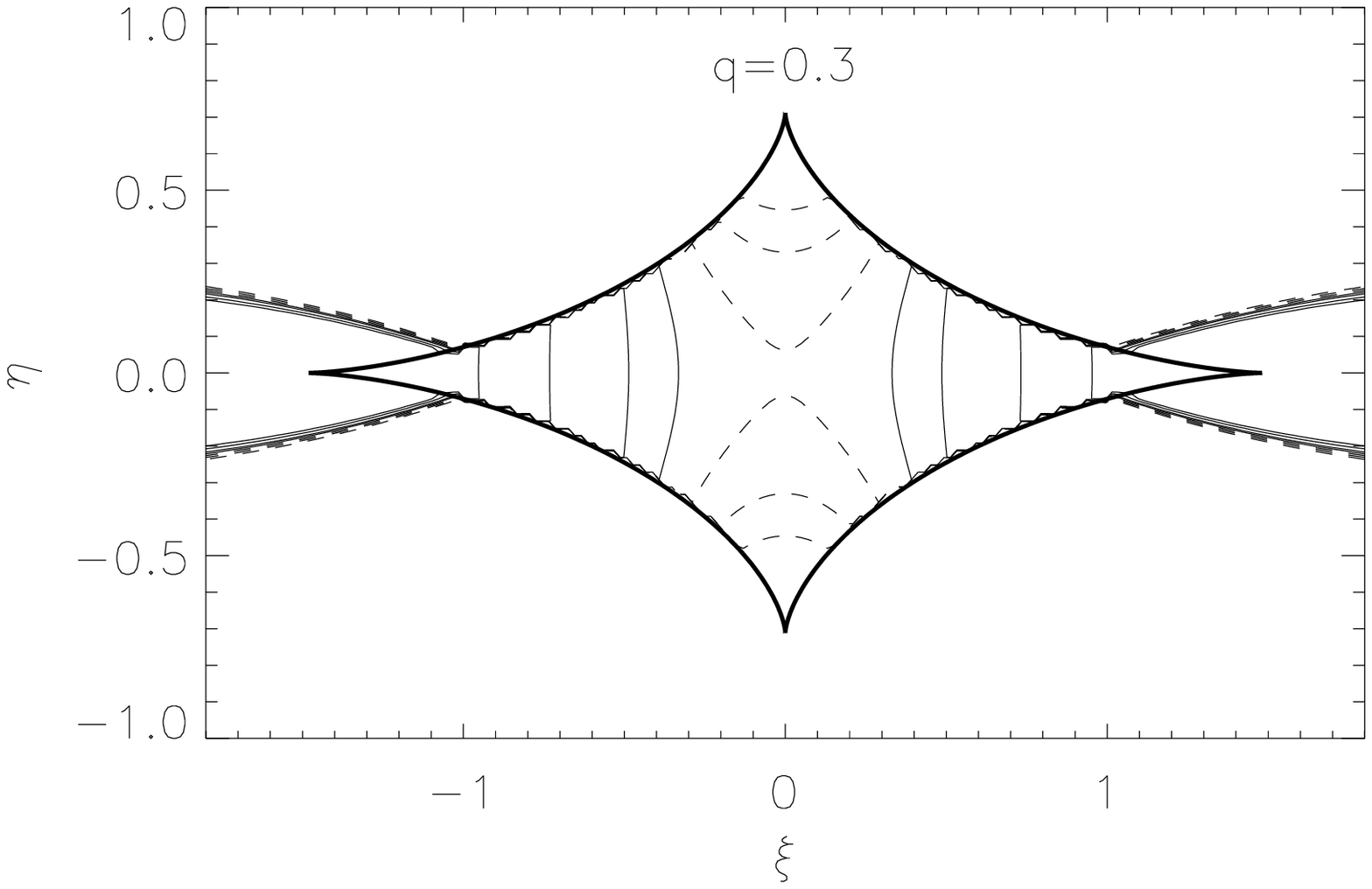}
\caption{
The {\it left} panel shows the contours for the sum of signed magnifications
for $q=0.9$.
The axes are in units of $a$ in eq. (\ref{sed}). 
The thick diamond-shaped curve is the caustic. The three dashed
contours (from outer to inner) have magnifications 2.785, 2.79, and 2.795
respectively. The three solid contours (from inner to outer)
have magnifications 2.8, 2.805 and 2.81, respectively.
The outgoing contours from the cusp are those for two image configurations.
The {\it right} panel is for $q=0.3$. The dashed
contour levels (from outer to inner) are 2.7, 2.725, and 2.75
respectively, while the four solid contour levels (from inner to outer) are
2.775, 2.8, 2.85, and 2.9, respectively.
The outgoing contours (closely spaced together)
from the cusp are contours
for two image configurations. Notice the scales on the
two axes are different.
}
\end{figure}

\subsection{The case with shear}

The lens equation with a shear is more complicated and can
be studied only numerically. However, we find that for most cases,
the sum of signed magnification is still around 2--3. In Fig. 2, we show
one example for $\gamma=0.1, \phis=45^\circ$ and $q=0.8$. For this
set of parameters, the caustic structure is a deformed diamond, which
divides the four image region (inside the diamond)
and the two image region (outside the diamond).
The sum of signed magnification at the origin is 2.47, while
most areas inside the caustics have magnifications between 2.2 to 2.8.

\begin{figure}
\epsfxsize=0.75\columnwidth \epsfbox{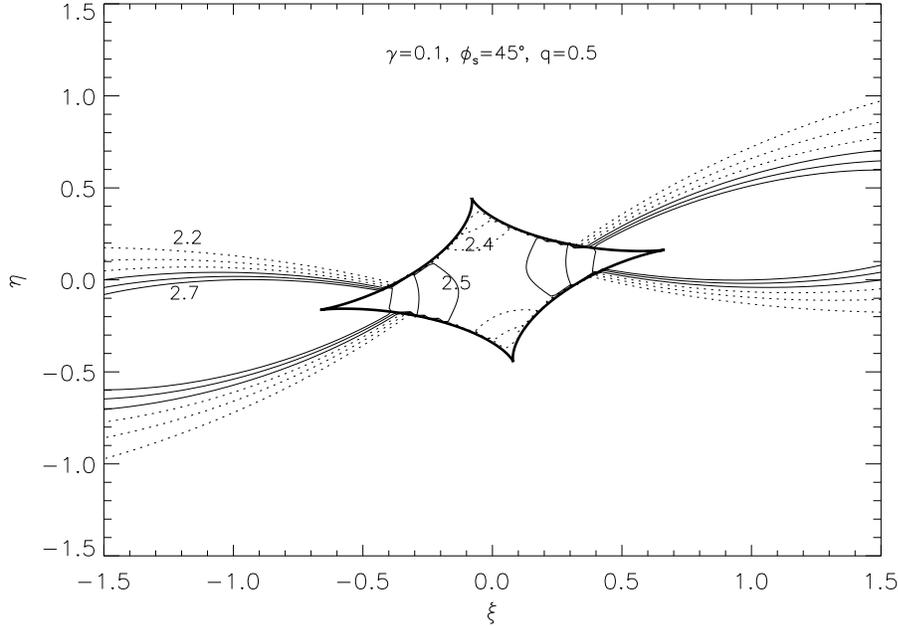}
\caption{
Contours for the sum of signed magnifications
for $q=0.9, \gamma=0.1$ and $\phis=45^\circ$.
The axes are in units of $a$ in eq. (\ref{sed}). 
The thick diamond-shaped curve is the caustics. The three dotted
contours have total magnifications 2.2, 2.3, and 2.4
respectively. The three solid contours
have magnifications 2.5, 2.6 and 2.7, respectively.
The outgoing lobes originating from the cusp are contours for two 
image configurations.
}
\end{figure}

However, there are significant exceptions to this. For 
some combinations of shear and ellipticity, the total signed
magnification falls significantly outside the range of 2-3. The reason 
for this is that qualitatively different image configurations
appear. More specifically, for some combinations of the $\gamma$ and $\eps$,
the caustics become small, and
higher-order singularities (such as swallowtails) emerge. Fig. 3 shows
how the caustics change for $q=0.6$ when the strength of 
shear changes but keep its
angle fixed at $\phis=90^\circ$. When the shear
is $\lesssim$ 0.1, the caustics form a diamond, similar to the case without
shear. When the shear approaches a critical value ($\approx 0.15$ for 
our case), swallowtails begin to develop,
first along the $y$ axis (see the top right panel),
then along the $x$ axis (lower right). For some intermediate shear value
($\gamma=0.165$, lower left), complex caustics appear. Finally,
when the shear becomes sufficiently large
($\gtrsim 0.21$), the caustics (not shown) again covert
to a simple diamond. For a source at the
origin, the maximum number of images is eight, as can be seen for the
case with $\gamma=0.165$ (lower left panel). For this case, the
total signed magnification the four image configurations spans a
considerable range, from $\approx 0-20$ for various 4-image caustic
regions. These values are
considerably different from 2 found for the isothermal potential, or
2.8 for the isothermal elliptical density distribution without
shear. The magnification
for the central region with 8 images is approximately 5.06;
this magnification does seem remarkably constant
within 1\% or so.

\begin{figure}
\epsfxsize=\columnwidth \epsfbox{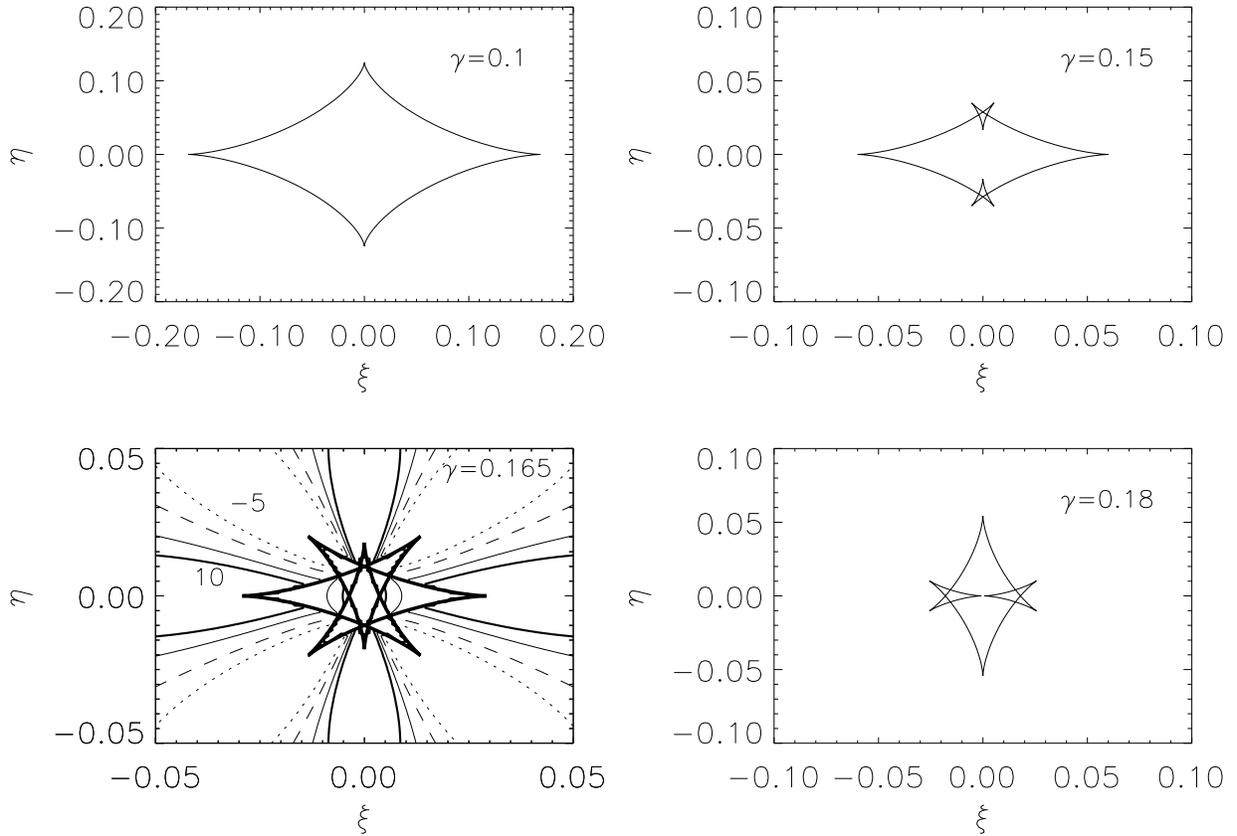}
\caption{
Caustic structures for an elliptical
density distribution with $q=0.6$ and $\phis=90^\circ$ as a function of
the magnitude of shear $\gamma$. 
The axes are in units of $a$ in eq. (\ref{sed}). 
The value of shear $\gamma$ is indicated in each
panel. Notice how swallowtail singularities appear and disappear as 
the shear changes. For $\gamma$ $\gtrsim$ 0.2, the caustics (not shown)
are again a diamond, similar to the one shown at the top left, only
elongated along the $y$-axis. For $\gamma=0.165$, the contours
for the total signed magnification
are shown for levels of 1.5 (dotted), 3 (dashed), 5.09 (thin solid),
and 8 (thick solid). The thick complex-shaped curve is the
caustic. 
} 
\end{figure}

We have also checked for higher moments of the image positions 
weighted by the signed magnifications for the elliptical
density distributions. We find that in general
these moments no longer satisfy the simple relations we presented in
\S 2, so we do not discuss these relations further.

\section{Summary and Discussion}

In this paper, we have studied the elliptical potentials using a new
moment approach. We have shown that
the image positions weighted by signed magnifications
satisfy fairly simple relations for this class of potentials. For
example, the total signed magnifications is 2 for an isothermal
elliptical potential, independent of the axial ratio (ellipticity);
similarly simple results hold for higher moments.
Our results confirm and extend those of Dalal (1998)
on the power-law elliptical potentials. These results provide
insights into the lensing properties of elliptical potentials. 
With improved image positions and flux ratios for lenses,
the moment approach should allow us to understand more easily whether
the power-law elliptical potentials can fit the observations data
at all, without much numerical effort.

However, we also find that these relations do not hold for its cousin --
the singular isothermal elliptical density distribution. For
this elliptical density
distribution without shear, the total signed magnification is 2.8,
again nearly independent of the ellipticity, but significantly different
from the value (2) for an isothermal potential.
For the case with shear, the magnification relations
become more complicated. For most cases we checked
the total signed magnification is
still around 2--3. However, the magnification no longer
changes with the shear {\it quadratically} as in eq. (\ref{mumu}),
instead it is much more sensitive. There are also
significant exceptions for the total signed magnification for some
combinations of shear and ellipticity. The most striking feature is that
6 or 8 image configurations can occur (see Fig. 3). Notice, however,
that the caustic structures are very small, so such image configurations
may have too small cross-sections to be seen
unless one observes a large number of
quasars. This is likely to be true even after one
takes into account the high magnification bias expected
for these cases. Such cases, if discovered, will be easy to identify
since the 6 or 8 images are located approximately on a circle.

The comparison between magnification relations in
elliptical density distribution and elliptical potentials highlights the
fact that magnifications are highly sensitive to the detailed form one
uses for the lensing potential,
much more so than the image positions; this is easily
understood intuitively
since magnifications involve higher order derivatives of the lensing potential
which tend to amplify small differences in the parameters. Since even the
more realistic elliptical density distribution is only an approximation
for the real lensing potential, as evidenced by the twisted isophotes
and many other irregular features in the light distribution, it is a
reasonable extrapolation that the predicted magnifications from simple
lens models
can be significantly different from the observed ones. Indeed, for some
radio lenses, one can show that even substructures with mass $\gtrsim
10^6 M_\odot$ can significantly influence the magnifications (Mao \&
Schneider 1998). It means that it will be difficult to model the flux
ratios using simple lens models. However, this also means that with the
steadily improving lens flux ratios and astrometries (e.g., see Blanton,
Turner \& Wambsganss 1998 for 2237+0305), we may be able to probe the
structures in lensing galaxies in great detail and
discriminate different classes of lensing potentials with confidence.

\section*{Acknowledgments}
We thank Peter Schneider for helpful discussions and comments
on the manuscript.
This work was partially supported by the ``Sonderforschungsbereich
375-95 f\"ur
Astro--Teil\-chen\-phy\-sik" der Deutschen For\-schungs\-ge\-mein\-schaft.

{}

\newpage

\bsp
\label{lastpage}
\end{document}